\documentclass[aps,pra,reprint,groupedaddress,longbibliography]{revtex4-2}
\usepackage{graphicx}
\usepackage{verbatim}
\usepackage{subfigure}
\usepackage[english]{babel}
\usepackage{siunitx}
\usepackage{pifont}
\usepackage{amsmath}
\usepackage{adjustbox}
\usepackage{wasysym}
\usepackage{lipsum}  
\usepackage{mathtools} 
\usepackage{upgreek}
\usepackage{multirow}
\usepackage{xcolor}
\usepackage{graphicx}
\usepackage{parskip}
\usepackage[margin=0.75in]{geometry}
\usepackage{dcolumn}
\usepackage{siunitx} 
\usepackage{booktabs} 

\begin{document}

\title{ Metrology in a two-electron atom: The ionization energy of metastable triplet helium ($\mathbf{2\,^3}S_\mathbf{1}$)   } 
\author{Gloria Clausen$^1$, Kai Gamlin$^{1,3}$, Josef A. Agner$^1$, Hansj\"urg Schmutz$^1$, Fr\'ed\'eric Merkt$^{1,2,3}$}

\affiliation{$^1$Department of Chemistry and Applied Biosciences, ETH Zurich, CH-8093 Zurich, Switzerland\\
	$^2$Quantum Center, ETH Zurich, CH-8093 Zurich, Switzerland\\
	$^3$Department of Physics, ETH Zurich, CH-8093 Zurich, Switzerland}

\date{\today}

\begin{abstract}
Helium (He) is the ideal atom to perform tests of ab-initio calculations in two-electron systems that consider all known effects, including quantum-electrodynamics and nuclear-size contributions. Recent state-of-the-art calculations and measurements of energy intervals involving the He $2\;^3S_1$ metastable state reveal discrepancies at the level of $7\,\sigma$ that require clarification both from the experimental and theoretical sides. 
We report on a new determination, with unprecedented accuracy, of the ionization energy $E_\mathrm{I}\,(2\;^3S_1)$ of the $(1s)(2s)\;^3S_1$ metastable state of He. The measurements rely on a new approach combining interferometric laser-alignment control, SI-traceable frequency calibration and imaging-assisted Doppler-free spectroscopy. With this approach we record spectra of the $np$ Rydberg series in a highly-collimated cold supersonic beam of metastable He generated by a cryogenic valve and an electric discharge. Extrapolation of the Rydberg series yields a new value of the ionization energy ($E_\mathrm{I}\,(2\;^3S_1)/h= 1\,152\,842\,742.7082(55)_\mathrm{stat}(25)_\mathrm{sys}\,\mathrm{MHz}$) that deviates by $9\,\sigma$ from the most precise theoretical result ($1\,152\,842\,742.231(52)\;\mathrm{MHz}$), reported by Patk\'o\v s, Yerokhin and Pachucki [Phys. Rev. A. \textbf{103}, 042809 (2021)], confirming earlier discrepancies between experiment and theory in this fundamental system.	
\end{abstract}

\maketitle

\section{Introduction}
As a fundamental three-body, two-electron system, the helium atom has played and still plays a central role in the development of quantum mechanics \citep{bethe57a,kabir57a,araki57a,schwartz61a, schwartz64a,douglas74a, drake85a,martin87a,drake94a, drake99a,morton06a, drake08a, pachucki06b, pachucki06c,pachucki17a, drake20a, yerokhin18a, patkos20a, patkos21b, patkos21a}. Over the years, calculations and measurements of the spectrum of He with ever increasing precision have led to a remarkably detailed understanding of the structure of this important atom. The theoretical framework needed to obtain precise level energies includes the treatment of relativistic and quantum-electrodynamics corrections and the evaluation of nuclear-size and nuclear-polarization effects. This
framework has been systematically developed in the last decades and has
been confirmed through comparison with experimental results. 
\\ \indent Spectroscopic measurements involving the $(1s)^2\;^1S_0$ ground state of He require short-wavelength radiation and do not yet reach the precision needed to test the calculations. Instead, almost all precision measurements in He involve one of the two metastable states resulting from the excited $(1s)(2s)$ configuration, the $2\;^1S_0$ and $2\;^3S_1$ states. These states can easily be generated at high densities in discharges and their very long lifetimes make them ideally suited as initial states of transitions to a broad range of final states with wavelengths ranging from the near-infrared to about 260 nm. He in the $2\;^3S_1$ state is of particular interest because it can be efficiently laser cooled \citep{aspect90a,morita91a, lawall95a, rooijakkers96a, pereiradossantos01a, vassen01a,koelemeij03a,keller14a, clausen24a} and manipulated by magnetic fields \citep{dulitz15a, chen20a,clausen24a}. The frequencies of numerous transitions involving states of different orbital-angular momenta and different spin multiplicities have been measured with high accuracy in both $^4$He \citep{sun20a, shiner94a,  rengelink18a,  luo13b, luo13a, canciopastor04a,  zheng17a, dorrer97a, luo16a, huang18a, clausen21a} and $^3$He \citep{rooij11a, vanderwerf23a, canciopastor12a, shiner95a} and the term values of many excited states of He are tabulated in the NIST atomic database \citep{nist_asd_template}. Although the overall agreement between theory and experiment can be regarded as impressive, reaching the level of $\Delta \nu/\nu < 10^{-10}$ for some transition frequencies, recent works also point out discrepancies and inconsistencies: 
\begin{itemize} \item Whereas the experimental frequencies of the $ (1s)(2p)\;^3P_J \leftarrow(1s)(2s)\;^3S_1$ transitions \citep{canciopastor04a, canciopastor12a, zheng17a} agree with theoretical values within the uncertainties, those of the $(1s)(3d)\;^3D_J \leftarrow(1s)(2s)\;^3S_1 $ \citep{dorrer97a} and $(1s)(3d)\;^3D_J \leftarrow (1s)(2p)\;^3P_J$ \citep{luo16a} transitions deviate from calculated values \citep{morton06a, wienczek19a, patkos21a} by 7 and 15 $\sigma$, respectively.
\item The experimental values of the ionization energies of the $(1s)(2s)\,^1S_0$ and $^3S_1$ states \citep{clausen21a} differ from the calculated values \citep{drake08a, pachucki17a, patkos21a} by $7\,\sigma$.
\item $^3$He/$^4$He isotopic shifts of several transitions yield inconsistent values of the nuclear charge radii; moreover, different measurements of the same transition frequencies do not always agree \citep{shiner95a, canciopastor12a, zheng17a, rooij11a, vanderwerf23a}. Very recent theoretical progress in the treatment of singlet-triplet mixing \citep{qi24a, pachucki24a} indicates that some of these inconsistencies may be resolvable.
\item The results of recent $n=2$ Lamb shift measurements in muonic $^3$He$^+$ and $^4$He$^+$ yield nuclear charge-radii differences that are inconsistent with those derived, with the help of theory, from measurements in ``nonexotic" helium \citep{vanderwerf23a, krauth21a, schuhmann23a}. Here also, at least some of the inconsistencies may be explained through improved treatment of singlet-triplet mixing \citep{qi24a, pachucki24a}.
\end{itemize}
These discrepancies and inconsistencies require attention because they may indicate limits of, and problems in, the theoretical framework. Large efforts are currently being invested in checking and improving theory and the calculations \citep{yerokhin22a,yerokhin23a, qi24a, pachucki24a}.  Measurements at improved precision and accuracy are also needed, not only to verify previous work, but also to provide benchmarks for future calculations. In this context, the use of different experimental approaches, subject to different systematic errors, is desirable. 
\\ \indent We report on a new measurement of the ionization frequency of the $2\;^3S_1$ metastable state of $^4$He at a 5-fold improved precision over the previous most accurate experimental determination of this quantity \citep{clausen21a}, which was obtained by adding the ionization frequency of the $2\;^1S_0$ state to the  
$(1s)(2s)\;^1S_0 \leftarrow (1s)(2s)\;^3S_1$ transition frequency \citep{rengelink18a}. The ionization energy of the $2\;^3S_1$ state is obtained here directly from the $np\;^3P_J\,\leftarrow2\,^3S_1$ Rydberg series based on a new experimental approach combining (i) supersonic beams of cold metastable He generated by a cryogenic valve equipped with a dielectric barrier discharge and delivering short ($20\,\mu$s duration) gas pulses; 
(ii) a new interferometric laser-alignment procedure to cancel 1$^\mathrm{st}$-order Doppler shifts that can be used at wavelengths below $\sim280$ nm, where optical fibers do not fulfill the desired specifications;
(iii) an imaging-assisted Doppler-free spectroscopic technique introduced recently \citep{clausen23a}, and (iv) an SI-traceable frequency-calibration procedure of the $261$ nm laser needed to record spectra of the $(1s)(np)\;^3P_{J=0-2}\,\leftarrow(1s)(2s)\,^3S_1$ transitions. 
A report on an earlier phase of the measurement, emphasizing our new imaging-assisted
technique to suppress the first-order Doppler shifts, has been presented in Ref. \onlinecite{clausen23a}. In
the present article, we give a full account of the measurements, including a detailed description of
the procedure we have established to evaluate and minimize the main systematic errors. To anticipate the main result of this investigation, we found that the $7\,\sigma$ discrepancy in the ionization energy of the $2\,^3S_1$ state of He has now increased to $9\,\sigma$, dominated by the uncertainty of the theoretical result.
\section{Experimental Setup}
\begin{figure*}[ht]\includegraphics[width=\textwidth]{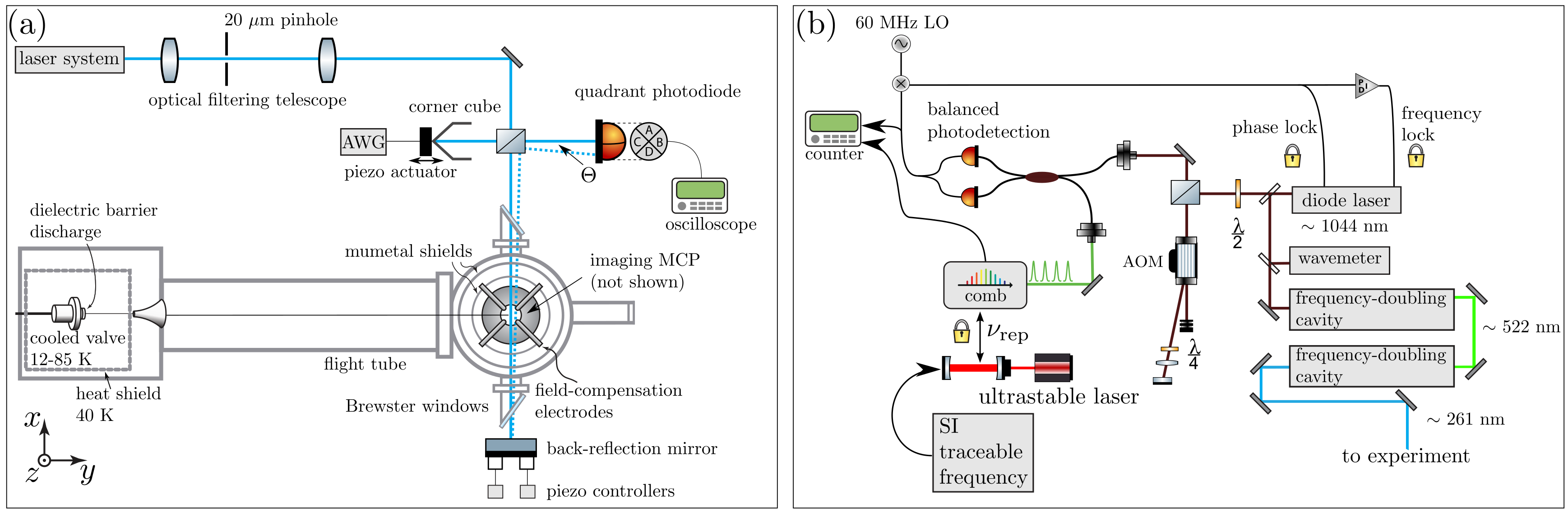}
\caption{\label{fig:1} Schematic overview of the experimental setup. (a) Bottom: Vacuum system with valve, discharge electrodes and skimmer used to generate the cold He$^*$ beam, the flight tube, and the photoexcitation chamber including two concentric cylindrical mu-metal shields and field-compensation electrodes, and the mirror used to retroreflect the laser beam. Top: Optical system with mode-cleaning optical-filtering telescope and interferometer used to align the incoming and retroreflected laser beams using a corner cube and a quadrant photodiode. (b) Laser system and frequency-calibration setup. Right: Tapered-amplified diode laser with the two frequency-doubling cavities used to generate cw-UV radiation at 261 nm. Center: Double-pass  acousto-optic-modulator (AOM) including cat-eye retroreflection setup. Left: Beat-detection unit used to phase lock the diode laser to the stabilized frequency comb (AWG: arbitrary waveform generator; MCP: microchannel-plate detector; LO: local oscillator). See text for details.  }
\end{figure*}

The experimental setup is illustrated schematically in Fig.~\ref{fig:1}(a). The experiment is carried out using a supersonic-beam apparatus under high-vacuum conditions. A supersonic beam of helium is generated in the source chamber by a cryogenically cooled pulsed valve that releases short (duration $\approx 20\,\mu$s) gas pulses at a repetition rate of 8.33 Hz. The valve temperature can be adjusted in the range from 12 to 85 K, leading to beam velocities between $480(15)\,\mathrm{m/s}$  and $1080(40)\,\mathrm{m/s}$, respectively, where the numbers
in parentheses give the respective full widths at half maximum of the velocity distributions. A small fraction of the helium atoms are excited to the metastable $(1s)(2s)\,^3S_1$ state (He* hereafter) in a dielectric-barrier discharge at the orifice of the nozzle. A 0.8-mm-diameter skimmer located 40 cm downstream from the nozzle is used to select the transversely cold He$^*$ atoms. After a 1.35-m-long field-free flight tube, the atoms enter the photoexcitation region, where He$^*$ is excited to $np$ Rydberg states using single-mode continuous-wave UV laser radiation ($\lambda \sim 261$ nm). The photoexcitation region is magnetically shielded by two concentric mumetal shields. Stray electric fields are compensated in three dimensions using a stack of electrodes \citep{osterwalder99a,beyer18a}. The transitions are detected by field ionization of the Rydberg states with a pulsed electric field of 0.75 kV/cm, which also extracts the He$^+$ ions toward a microchannel-plate detector (MCP) in the direction perpendicular to the plane defined by the laser and supersonic beams. Doppler-free spectra of the $(1s)(np)\,^3P_J\,\leftarrow\,(1s)(2s)\,^3S_1$ transitions are recorded using Imaging-Assisted Single-photon Spectroscopy (IASS hereafter) \cite{clausen23a}. The ionization energy of the $(1s)(2s)\,^3S_1$ state is determined by extrapolation of the $np$ Rydberg series to their limits. \\ \indent Detailed descriptions of the vacuum system, the supersonic beam, and the laser system have been provided in an earlier publication \cite{clausen23a}. We focus here on the description of the interferometer used to eliminate the residual 1$^\mathrm{st}$-order Doppler shift, on the frequency-calibration system, and on the discussion of the systematic uncertainties resulting from the measurement procedure.
\subsection{Laser system}
The UV laser system used to record spectra of the $(1s)(np)\,^3P_J\,\leftarrow\,(1s)(2s)\,^3S_1$ transitions is displayed schematically in Fig. \ref{fig:1}(b). To generate UV radiation tunable in the range $\lambda_\mathrm{UV}=260\, \mbox{--}\,265$ nm with output powers between 20 and 100 mW, the near-infrared (NIR) continuous-wave (cw) output of a tapered-amplified diode laser with wavelengths $\lambda_\mathrm{NIR}$ tunable in the range $\lambda_\mathrm{NIR}=1040\, \mbox{--}\,1060$ nm is frequency quadrupled using two successive frequency-doubling cavities (\textit{Toptica, TA FHG}). A mode-filtering collimating telescope, consisting of a 20-$\mathrm{\mu m}$-diameter pinhole located in the focal planes of a 10-cm-focal-length lens on the incoming side and a 20-cm-focal-length recollimating lens on the outgoing side, is used to select the TEM$_{00}$ mode of the UV-laser beam and to enlarge the $1/e^2$ beam-waist diameter (in intensity) to about 2 mm (see Fig. \ref{fig:1}(a) top). \\ \indent
\begin{figure}[ht]\includegraphics[width=\columnwidth]{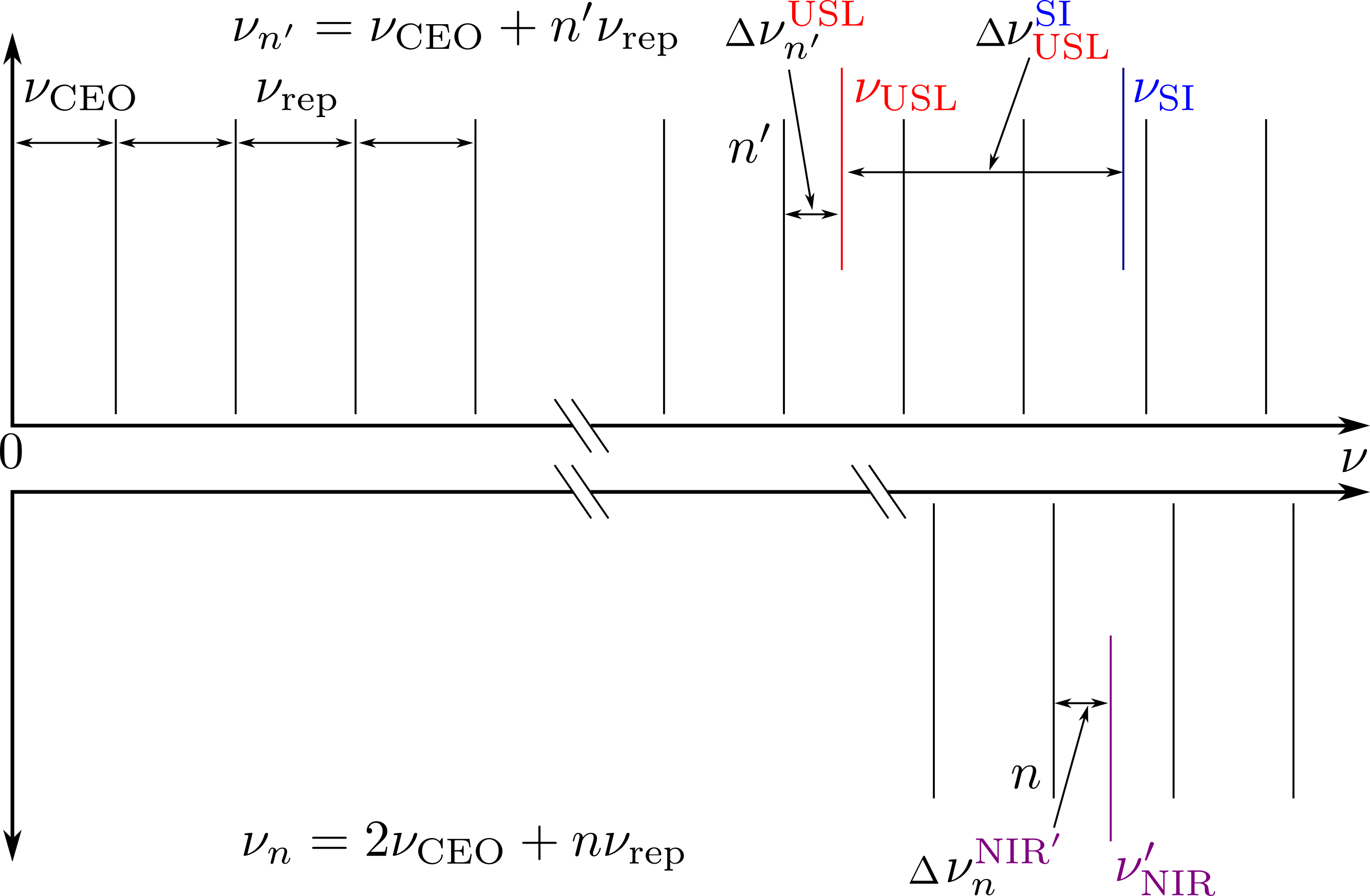}
\caption{Schematic illustration of the frequency-calibration procedure using the fundamental (upward) and frequency-doubled (downward, inverted scale) outputs of a frequency comb with carrier-envelope offset $\nu_\mathrm{CEO}$ and repetition rate $\nu_\mathrm{rep}$. The comb teeth are indicated by equidistant black lines. The repetition rate is stabilized by an ultrastable laser at a frequency $\nu_\mathrm{USL}$ (red line) using a lock to the $n'$th comb tooth. SI traceability is ensured through referencing the ultrastable laser to $\nu_\mathrm{SI}$ (blue line). The frequency-shifted fundamental output of the spectroscopy laser $\nu'_\mathrm{NIR}$ is phase locked to the $n$th tooth of the frequency-doubled output. See text for details. \label{fig:comb} }
\end{figure}  
$1\,\%$ of the fundamental output of the diode laser is used for frequency calibration and linewidth stabilization following the procedure described in Ref. \onlinecite{scheidegger23a}. Half of this reflection is sent to a wavemeter (\textit{HighFinesse WS 7-60}) to calibrate the laser frequency at a precision of $<\,60\,\mathrm{MHz}$. The other half is focused into an acousto-optic modulator (AOM) (\textit{Pegasus MT200}) operated at a modulation frequency $\nu_\mathrm{AOM}$ between 150 and 250 MHz. The first-order diffraction of the AOM passes through a pinhole and is retroreflected into the AOM employing a mirror in a cat-eye configuration \citep{donley05a} after rotating the polarization by $90\,^\circ$. This double-pass AOM arrangement shifts the laser frequency by  $-2\nu_\mathrm{AOM}$ ( $\nu'_\mathrm{NIR}=\nu_\mathrm{NIR}-2\nu_\mathrm{AOM}$) and minimizes efficiency losses caused by laser misalignment. The reflected, frequency-shifted and polarization-rotated radiation is separated by a polarizing beam splitter and coupled into a single-mode polarization-maintaining fiber connected to an ultrastable frequency comb (\textit{MenloSystems, FC1500-ULN}). 
The repetition rate of the frequency comb (near 250 MHz) \begin{equation}
\nu_\mathrm{rep}=\frac{1}{n'}(\nu_\mathrm{USL}-\Delta\nu_{n'}^\mathrm{USL}-\nu_\mathrm{CEO}) \end{equation} is stabilized by an optical lock (beat frequency $\Delta \nu_{n'}^\mathrm{USL}$) between a selected comb tooth with tooth number $n'$ and frequency $\nu_n'=\nu_\mathrm{CEO}+n'\nu_\mathrm{rep}$  ($\nu_\mathrm{CEO}$ is the carrier-envelope offset frequency) and an ultrastable laser with frequency $\nu_\mathrm{USL}$ locked to a high-finesse cavity, see Fig. \ref{fig:comb}.
 Frequency drifts of the cavity are quantified by referencing the ultrastable laser to an SI-traceable frequency standard ($\nu_\mathrm{SI}$ with beat frequency $\Delta\nu_\mathrm{USL}^\mathrm{SI}$, see Fig. \ref{fig:comb}) provided by the Swiss Federal Institute of Metrology, METAS, through a stabilized fiber network \citep{husmann21a}. 
\\ \indent The frequency-comb output is frequency doubled, amplified and spectrally broadened in a \textit{MenloSystems M-VIS} unit consisting of an amplifier and  a photonic crystal fiber. The carrier-envelope
offset frequency is determined from the beat between frequency teeth of the
fundamental and frequency-doubled comb outputs. The comb teeth positioned around $\nu'_\mathrm{NIR}$ (tooth number $n$) are selected using a fine-grid blazed grating (\textit{Thorlabs GR13-1210}) and merged with the frequency-shifted diode-laser beam using a fiber combiner. The beat ($\Delta\nu_n^\mathrm{NIR'}$) between the frequency-shifted diode laser and the frequency comb is detected using a fiber-coupled balanced-photodetection unit (\textit{Thorlabs PDB425C-AC}). About half of the beat signal is used for frequency counting with a \textit{K+K Messtechnik FXM50} counter, while the other half is mixed with a local oscillator (LO) signal with frequency $\nu_\mathrm{LO}=60\,\mathrm{MHz}$ in a phase-frequency detector (\textit{Analog Devices, HMC403}) to generate an error signal. This error signal is used to control the current driving the diode laser and to phase lock the diode laser to the selected tooth of the frequency comb with a frequency offset of $\Delta\nu_n^\mathrm{NIR'}+2\nu_\mathrm{AOM}$. Slow frequency drifts are compensated with a proportional-integral-derivative lock provided by the \textit{Toptica Topas Lock Wizard} software. The stability of   $\nu_\mathrm{NIR}$ is determined in an in-loop measurement to be well below 1 kHz from the linewidth of the beat signal $\Delta \nu_n^\mathrm{NIR'}$ between the comb and the laser frequencies.
\\ \indent With this arrangement, the UV laser frequency $\nu_\mathrm{UV}=4\nu_\mathrm{NIR}=4(\nu'_\mathrm{NIR}+2\nu_\mathrm{AOM})$ is given by \begin{align}
\nu_\mathrm{UV}=&4\times \big [n\cdot\overbrace{\frac{1}{n'}(\underbrace{\nu_\mathrm{SI}-\Delta\nu_\mathrm{USL}^\mathrm{SI}}_{\nu_\mathrm{USL}}-\Delta\nu_{n'}^\mathrm{USL}-\nu_\mathrm{CEO})}^{\nu_\mathrm{rep}}\\&+2\nu_\mathrm{CEO}+\Delta\nu_n^\mathrm{NIR'}+2\nu_\mathrm{AOM}\big ]. \nonumber
\end{align} The comb-tooth numbers $n$ and $n'$ are determined in reference measurements using the wavemeter, exploiting the fact that half the comb mode spacing $(\nu_\mathrm{rep}/2)$ is larger than the wavemeter precision of $60$ MHz.
\subsection{1$^\mathrm{\textbf{st}}$-order-Doppler-shift suppression \label{sec:doppler}}
\begin{figure}[t]
\includegraphics[width=\columnwidth]{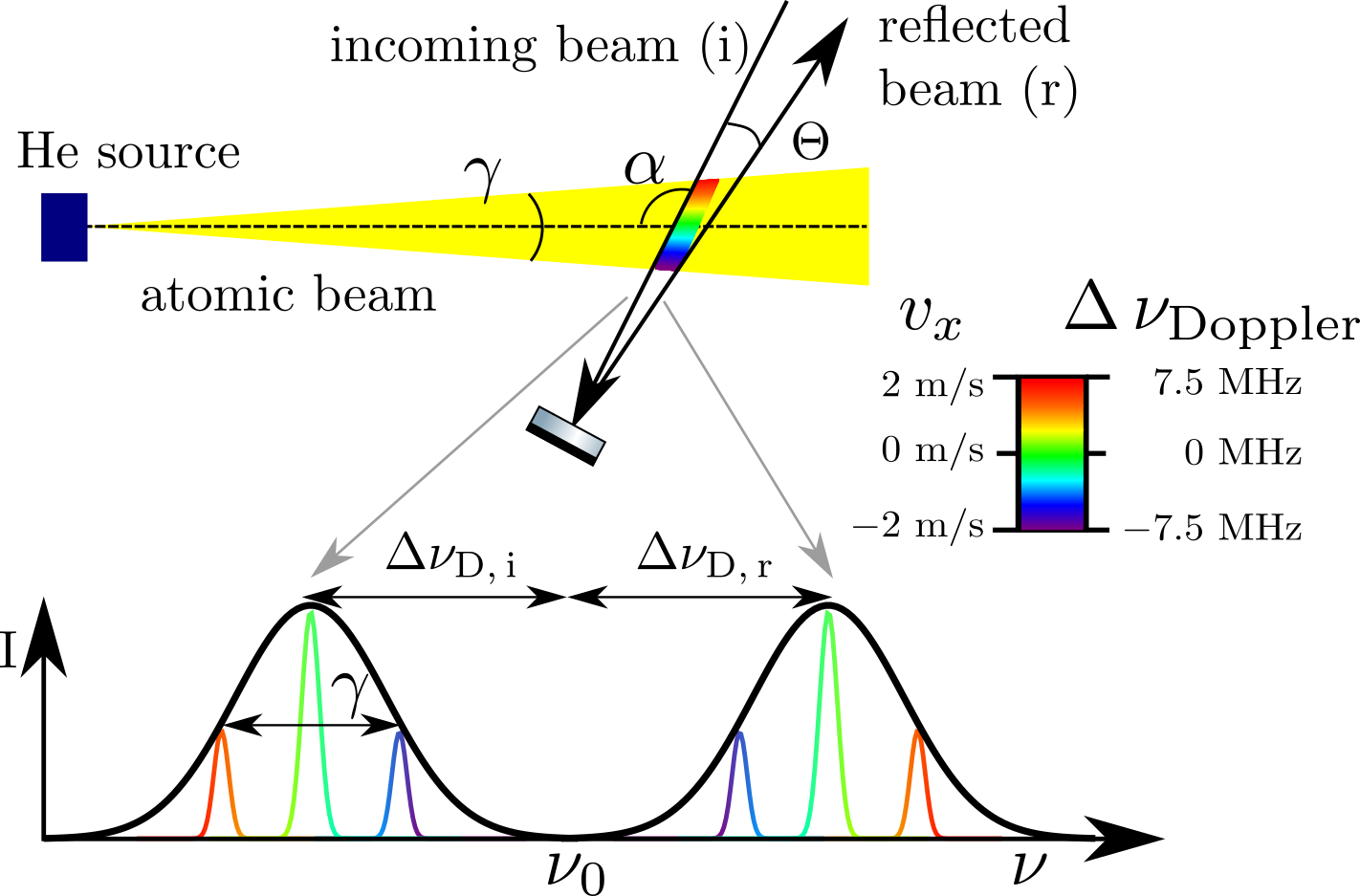}  
\caption{Principle of IASS, adapted from Ref.~\cite{clausen23a}. Top: 
Geometric arrangement of gas and laser beams with the relevant angles $\alpha,\,\gamma$ and $ \Theta$. The rainbow symbolizes the distribution of Doppler shifts. Bottom: Schematic spectrum with the two Doppler components corresponding to the incoming (i) and reflected (r) laser beams and the narrower components (colored lines) selected by imaging. See text for details. 
\label{fig:2}}
\end{figure}

The 1$^\mathrm{st}$-order Doppler shift $\Delta \nu_\mathrm{D}$ of a transition in an atom moving with velocity $\vec{v}$ intersecting a laser beam with a wave vector $\vec{k_\mathrm{L}}$ at an angle $\alpha$ is given by \begin{equation}
\Delta \nu_\mathrm{D}=\frac{\vec{k_\mathrm{L}}\cdot \vec{v}}{2\pi}=\frac{v\cdot \cos \alpha}{\lambda_\mathrm{L}}=\frac{v}{c}\cdot \nu_\mathrm{L}\cos \alpha.
\end{equation}
Although the experimental configuration guarantees an $\alpha$ value close to $90\,^\circ$, in practice small deviations from $90\,^\circ$ are unavoidable and the residual $1^\mathrm{st}$-order Doppler shift needs to be compensated. To this end, we retroreflect the laser beam with a mirror located beyond the photoexcitation region, ensuring that the incoming and retroreflected beams overlap, see Fig. \ref{fig:2}. This configuration leads to the observation of two Doppler components with opposite $1^\mathrm{st}$-order Doppler shifts, allowing the $1^\mathrm{st}$-order-Doppler-free frequency to be determined as the geometric center of the two components. \\ \indent Because of the correlation between the transverse velocity and the transverse position of the atoms in a supersonic beam one can, by imaging, record spectra with reduced Doppler width, as demonstrated in Ref. \onlinecite{clausen23a}. This is achieved by field ionization of the He $(1s)(np)\,^3P_J$ atoms with a pulsed electric field and extracting the He$^+$ ions in the direction perpendicular to the plane $(xy)$ defined by the laser and supersonic beams and monitoring the impact positions on the MCP-detector surface. Integrating the He$^+$-ion signal in distinct narrow parallel stripes of the images enables one to divide the Doppler-broadened line profiles into numerous narrower components, as represented schematically by the colored spectral lines in Fig. \ref{fig:2} (see also Fig. \ref{fig:6} below). To
optimize the spectral resolution, the widths of these stripes are narrowed down until the
linewidths no longer decrease.\\ \indent 
For a small deviation angle $\Theta$ from exact retroreflection, the $1^\mathrm{st}$-order Doppler shifts of the lines observed with the incoming and retroreflected laser beams are 
\begin{subequations}
\begin{align}
\Delta \nu_\mathrm{D, \,i}&=\frac{v}{c}\cdot \nu_\mathrm{L}\cos \alpha \label{eq:incident}  \\ \intertext{and}
\Delta \nu_\mathrm{D, \,r}&=-\frac{v}{c}\cdot \nu_\mathrm{L}\cos (\alpha+\Theta), \label{eq:reflected} 
\end{align} \end{subequations}respectively, which introduces a systematic error  \begin{equation}
\delta \nu_\mathrm{D}\approx \pm\frac{v}{2c}\nu_\mathrm{L} \Theta \label{eq:residual}
\end{equation} in the determination of the  $1^\mathrm{st}$-order-Doppler-free transition frequency. \\ \indent 
Standard methods to precisely retroreflect a laser beam and achieve conditions of perfect wavefront-retracing $180^\circ$ reflection include the recoupling of the reflected beam into the fiber out of which the forward-propagating beam emerged \citep{beyer16c, wirthl21a} or into another fiber \citep{wen23a}, and the direct reflection of the beam using prisms and corner cubes \citep{shiner94a,berkeland95a, brown13a}. Because our experiments require radiation with wavelengths below $\sim261\;\mathrm{nm}$, the former methods cannot be used due to the lack of adequate fibers. The latter method is accompanied by undesirable polarization changes and intensity losses. Instead, we modify an interferometric alignment procedure originally introduced by \citet{mueller05a} to align two beams of slightly different frequencies with a corner cube in the reference arm of the interferometer. We show below how this method can be extended to the retroreflection of a single-frequency beam. 
\\ \indent 
The alignment setup is depicted schematically in Fig. \ref{fig:1}(a) and consists of a Michelson interferometer equipped with a corner cube [\textit{Newport UBBR1}, certified alignment error $<1\,\mathrm{arcsec}$ ($\sim5\,\mu \mathrm{rad}$)] as retroreflector in the reference arm. After the mode-filtering telescope, the TEM$_{00}$ mode of the UV laser beam (beam diameter $d\approx 2\,\mathrm{mm}$) is guided to the 50:50 beam splitter of the Michelson interferometer. In the reference arm, the corner cube is mounted on a piezo ring chip, to which arbitrary potential waveforms can be applied to displace its position in the $y$ direction and thereby adjust the reference-path length. In the measurement arm, the laser beam enters the vacuum chamber through a Brewster window, traverses the photoexcitation region, and exits the vacuum chamber through a second Brewster window before being retroreflected by a highly reflective mirror. The two mirror-alignment axes are controlled with piezo elements [see Fig. \ref{fig:1}(a)]. After retroreflection, the two beams are recombined at the 50:50 beam splitter and the alignment is monitored in the interference arm by observing the interferogram using a quadrant photodiode (\textit{Hamamatsu S4349}). \\ \indent The misalignment angle $\Theta$ is encoded in the distance between the interference fringes and their orientation. 
 \begin{figure}[h]
\includegraphics[width=\linewidth]{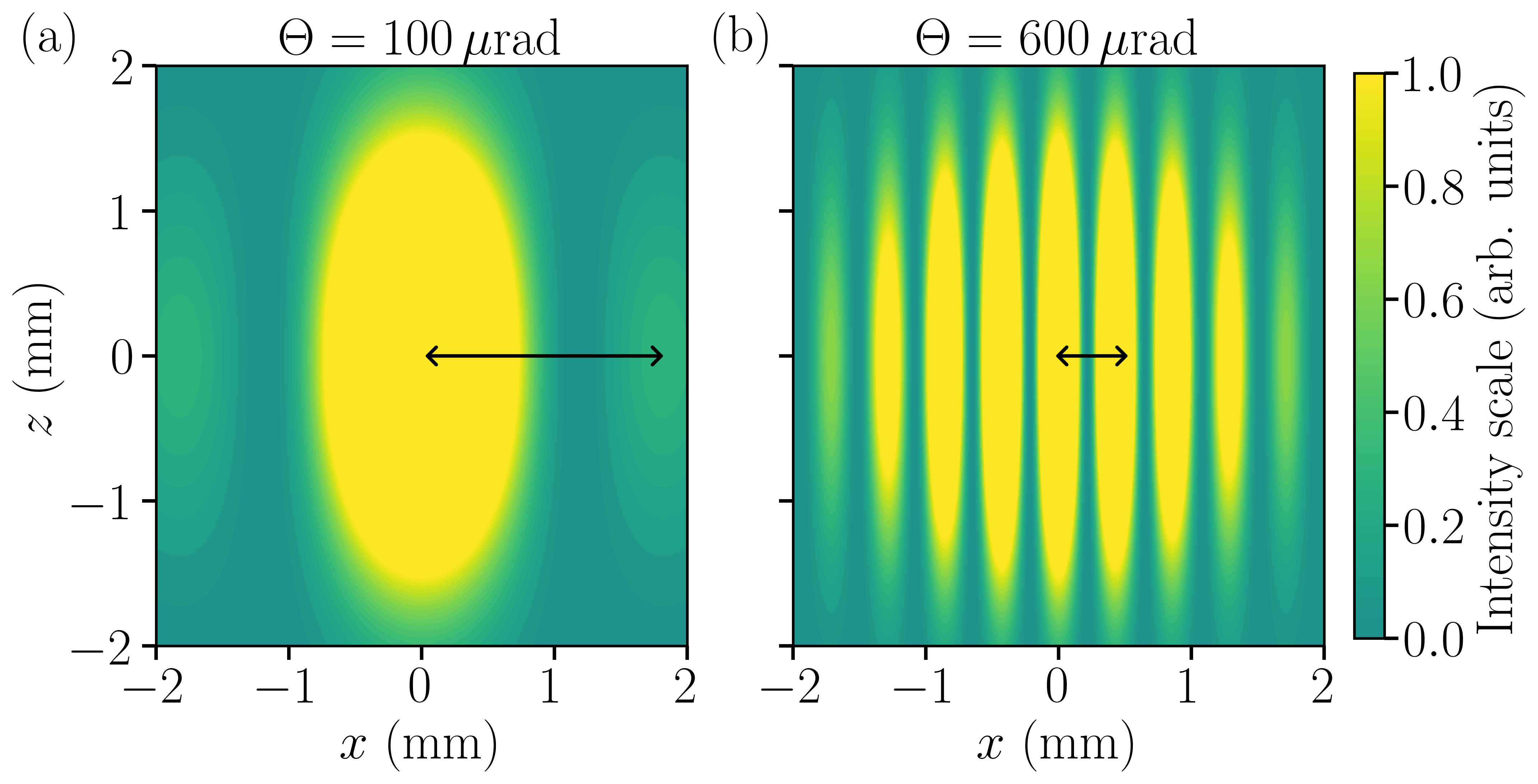}  
\caption{Calculated intensity patterns of the interferograms from reference and measurement laser beams intersecting at angles of $\Theta=100\,\mu\mathrm{rad}$ (a) and $\Theta=600\,\mu\mathrm{rad}$ (b). The fringe distance is indicated by the black arrows. \label{fig:3}}
\end{figure}
As illustration, Figs. \ref{fig:3} (a and b) display the interference patterns calculated for two Gaussian laser beams intersecting in the $xz$ plane at an angle $\Theta$ of $100\,\mu \mathrm{rad}$ and $600\,\mu \mathrm{rad}$, respectively. The first laser beam propagates along the $y$ axis while, for illustrative purposes, the second (measurement) laser beam is chosen to propagate in the $xy$ plane. In this configuration, the fringes are aligned along the $z$ direction and their distance is inversely proportional to $\Theta$. Using the interferogram of the two beams as alignment diagnostics, $\Theta$ can be reduced to below $100\,\mu\mathrm{rad}$ by maximizing the fringe distance in the observed interferogram. \\ \indent
\begin{figure}[ht]
\includegraphics[width=0.99\columnwidth]{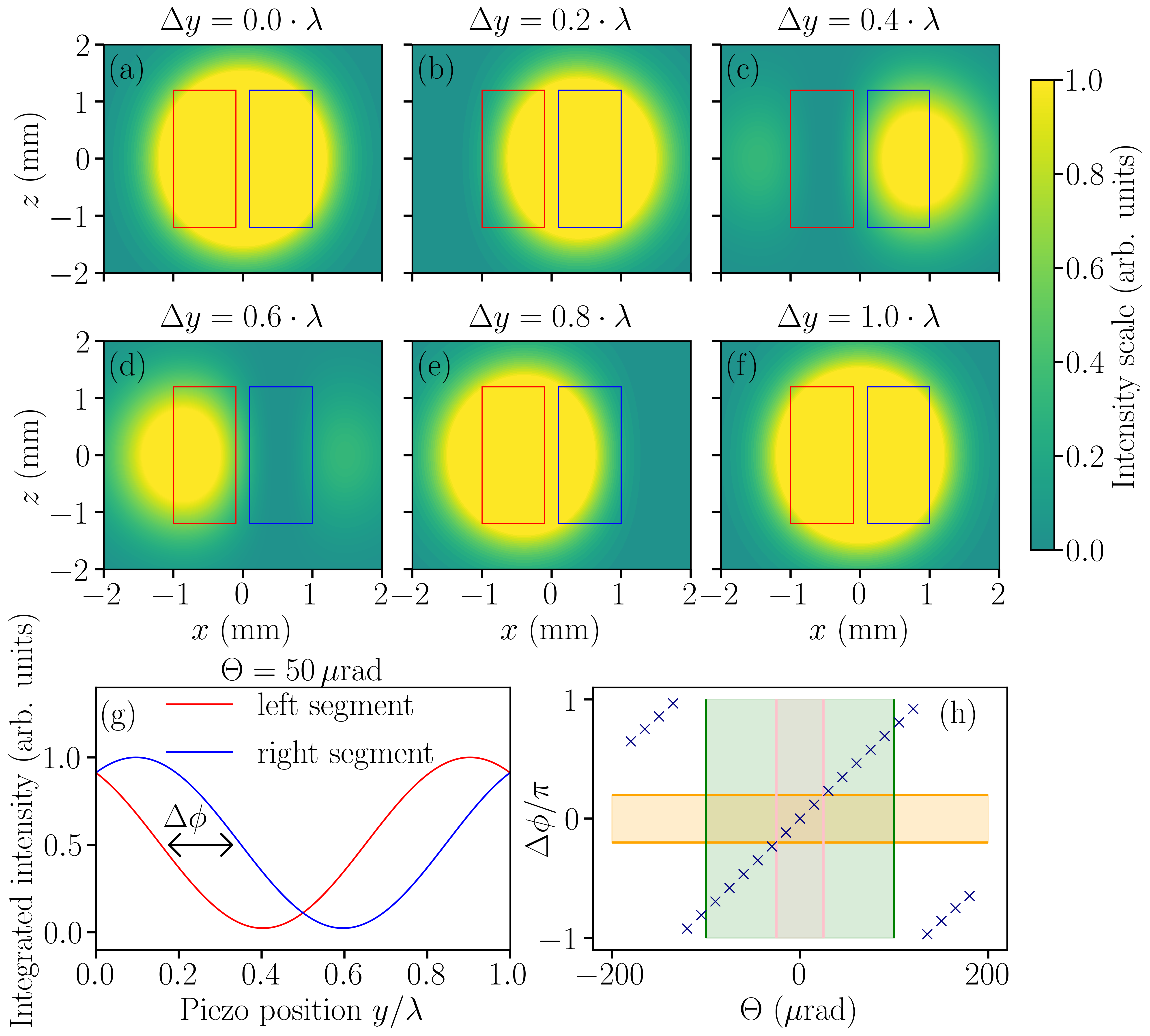} 
\caption{(a)-(f) Calculated interference patterns formed by two laser beams intersecting at an angle $\Theta=50\,\mu\mathrm{rad}$ for different reference-arm lengths of the interferometer. The left and right areas used for the intensity readout are indicated by red and blue frames, respectively. (g) Integrated intensities in the left (red) and the right (blue) areas of the interferograms as a function of the corner cube $y$ position. (h) Phase shift $\Delta \phi$ (blue crosses) of the intensity monitored in the blue and red areas of the quadrant photodiode as a function of the misalignment angle $\Theta$. The determination of the phase shift to better than $0.2\pi$ (orange) enables one to reduce the misalignment angle to less than $20\,\mu\mathrm{rad}$ (pink) if the coarse alignment of the mirror can be made with $100\,\mu\mathrm{rad}$ accuracy (green). At larger misalignment angles, the phase lag exhibits discrete jumps by integer multiples of $2\pi$.\label{fig:4}}
\end{figure} 
An intensity-fluctuation-independent readout of $\Theta$ is achieved by displacing the corner cube along the $y$ axis using the piezo ring chip.
The corresponding modification of the path length in the reference arm shifts the phase of the reference laser beam in the interference plane with respect to the measurement laser beam and alters the resulting interferogram. Figures \ref{fig:4}(a)-(f) illustrate the calculated changes in the interference patterns of two Gaussian beams intersecting at an angle $\Theta=50\,\mu\mathrm{rad}$, as the reference-path length is varied by up to $\lambda_\mathrm{UV}$. This variation results in a drift of the intensity maximum along the $x$ direction.  
Comparing the phase lag $\Delta \phi$ of the integrated intensities in the two regions of the $xz$ plane marked by red and blue frames in the interferogram (and corresponding to adjacent quadrants of the detector) while moving the cornercube $y$ position enables a direct readout of $\Theta$, as illustrated in \ref{fig:4}(g). As expected, a path-length shift of $\lambda_\mathrm{UV}$ causes a total phase change of the reference laser beam by $2\pi$. \\ \indent
The dependence of the calculated phase shift $\Delta \phi$ on the misalignment angle $\Theta$ is represented by the blue crosses in Fig. \ref{fig:4}(h). A fine alignment of the retroreflection mirror can therefore be achieved by minimizing $\Delta\phi$. Under our experimental conditions, pressure and temperature fluctuations, as well as slow drifts of optical components, limit the sensitivity of detection of phase lags $|\Delta \phi|$ to $0.2\pi$ rad.
 An accuracy of better than $\pm\,20\,\mu\mathrm{rad}$ can therefore be achieved in the alignment of the retroreflection, corresponding to systematic uncertainties in the residual $1^\mathrm{st}$-order Doppler shift of $\sim \pm 20$ kHz at a wavelength of 261 nm and a beam velocity of 500 m/s, see Eq. \ref{eq:residual}. \\ \indent 
To determine the residual misalignment angle $\Theta$ at an even higher accuracy than the $\pm\,20\,\mu\mathrm{rad}$ uncertainty guaranteed by the interferometric alignment procedure just described, we measure, for each alignment, the transition frequencies for different atomic-beam mean velocities between 500 and 1100 m/s, set by adjusting the valve temperatures between 10 and 85 K, and extrapolate the Doppler-free transition frequencies, see Sec. \ref{sec:results}. By repeating such measurements for several independent interferometric alignments and averaging the values of the extrapolated Doppler-free transition frequencies, the systematic uncertainties from the residual Doppler shifts are converted into a statistical uncertainty.
\section{Results \label{sec:results}}
\begin{figure}[ht]
\includegraphics[width=\linewidth]{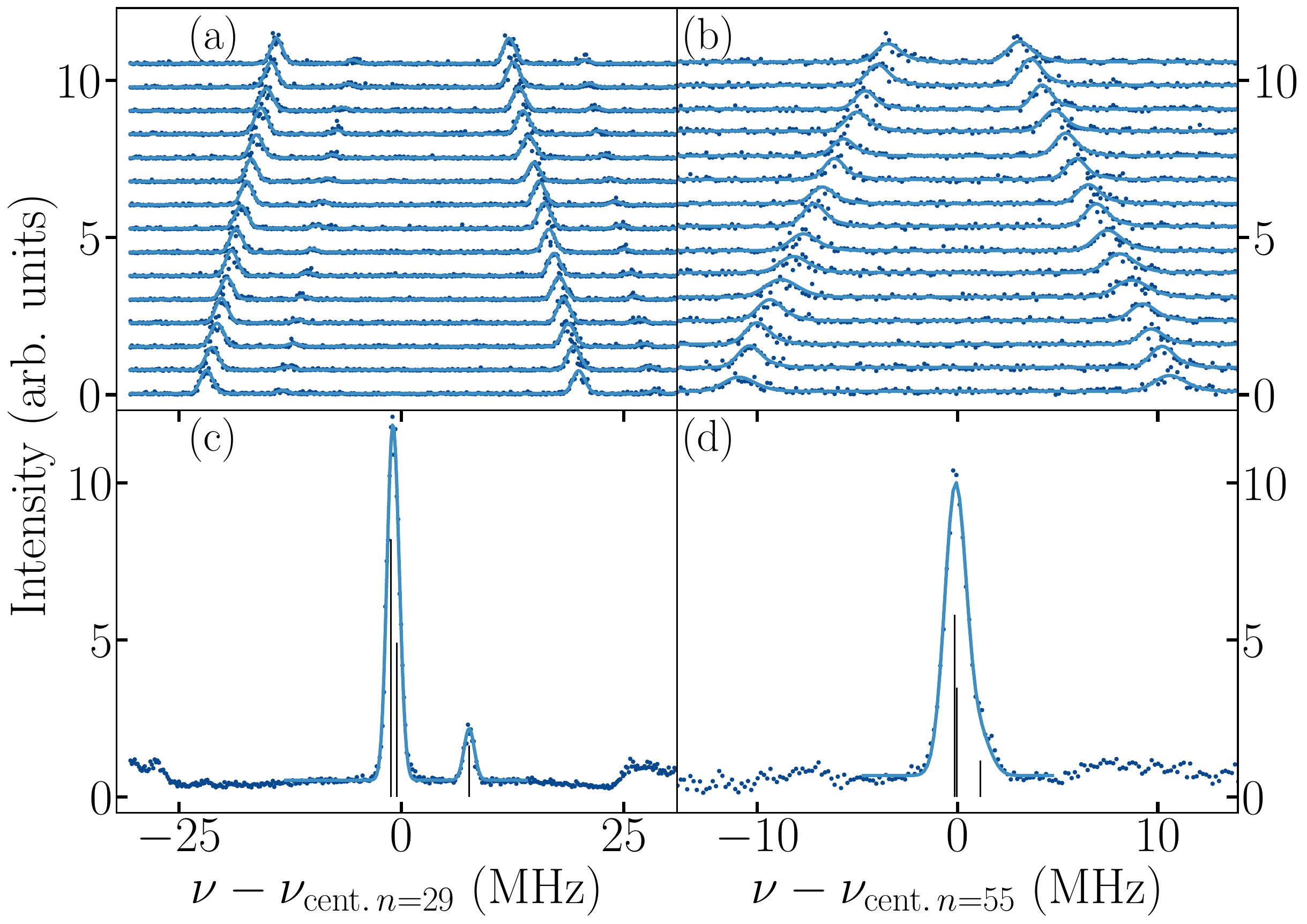} 
\caption{Imaging-assisted Doppler-free spectra, shifted along the intensity axis, of the (a) $(1s)(29p)\,^3P_J\,\leftarrow\,(1s)(2s)\,^3S_1$ and (b) $(1s)(55p)\,^3P_J\,\leftarrow\,(1s)(2s)\,^3S_1$ transitions in He, obtained by integration over neighboring stripes of the images and corresponding to different transverse-velocity components of the atomic beam (dark-blue dots) and fits to the experimental spectra (solid blue lines). (c) and (d): Sums of the corresponding cross-correlation spectra. The stick spectra give the positions of the $J=2,\,1$ and 0 fine-structure components in order of increasing frequency and the full blue line represents the spectrum calculated using the line-shape parameters determined in the least-squares fits. \label{fig:6}}
\end{figure}

To determine the ionization energy of metastable He$^*$, we measured transitions to the $(1s)(np)\,^3P_J$ Rydberg states with principal quantum numbers $n=27,\,29,\,33,\,35,\,40,\,50$, and 55. As illustrative examples, spectra of the $(1s)(29p)\,^3P_{J=0,1,2}\,\leftarrow\,(1s)(2s)\,^3S_1$ and $(1s)(55p)\,^3P_{J=0,1,2}\,\leftarrow\,(1s)(2s)\,^3S_1$ transitions are shown in Figs. \ref{fig:6}(a) and (b), respectively. The figures display the imaging-assisted Doppler-free spectra recorded by integrating the He$^+$ ion signal in parallel stripes of the detector surface, corresponding to different transverse-velocity components of the atomic beam. 
\\ \indent
At each $n$ value, the individual spectra can be combined into a single Doppler-free spectrum of much higher signal-to-noise ratio [see Figs. \ref{fig:6}(c) and (d)] using a circular-cross-correlation procedure \citep{clausen23a}.
The spectral lineshapes of the cross-correlation spectra are well described by the sum of three Gaussian profiles, corresponding to transitions to the fine-structure components $J=2,\,1$, and 0 of the $np\,^3P_J$ Rydberg states, with relative intensities of $5:3:1$, respectively. The spectral linewidths in Fig. \ref{fig:6} are primarily determined from the finite size of the valve and skimmer orifices and
to a lesser extent by the transit time of the atoms through the laser beam and by imperfections of the
ion-extraction electrode systems. The lifetimes of all states involved in the measured transitions are too long to make significant contributions to the linewidths. \\ \indent The fine-structure splittings between the $J=0,\,1$ and $2$ levels scale with $n^{-3}$. At $n=29,$ the $J=0$ component is well separated from the $J=1,\,2$ components, which are not resolved spectrally but lead to a broadened asymmetric line shape of the lower-frequency line. At $n=55$, one single line is observed, but the $J=0$ component can be clearly identified as a shoulder on its high-frequency side. The observed fine-structure splittings are perfectly described using the Rydberg-Ritz formula (see Eq. \ref{eq:rydberg} below) with the quantum defect parameters reported by \citet{drake99a}, and are used to determine the Doppler-free centroid frequencies in a least-squares-fit procedure.
The fit parameters used to describe the spectral lineshapes are the $1^\mathrm{st}$-order-Doppler-free centroid position, the amplitude of the $J=2$ component, a common linewidth for all fine-structure components, and a constant background offset. In the least-squares fits, the data points are weighted by the inverse signal intensity to account for the Poissonian nature of the ion-detection statistics. 
The positions of the fine-structure components used in the fit are given by the vertical sticks in Figs. \ref{fig:6}(c) and (d) and the blue traces represent the corresponding calculated spectra, which perfectly describe the observed intensity distributions.
\subsection{Doppler-free spectroscopy}
\begin{figure}[hb]
\includegraphics[width=\linewidth]{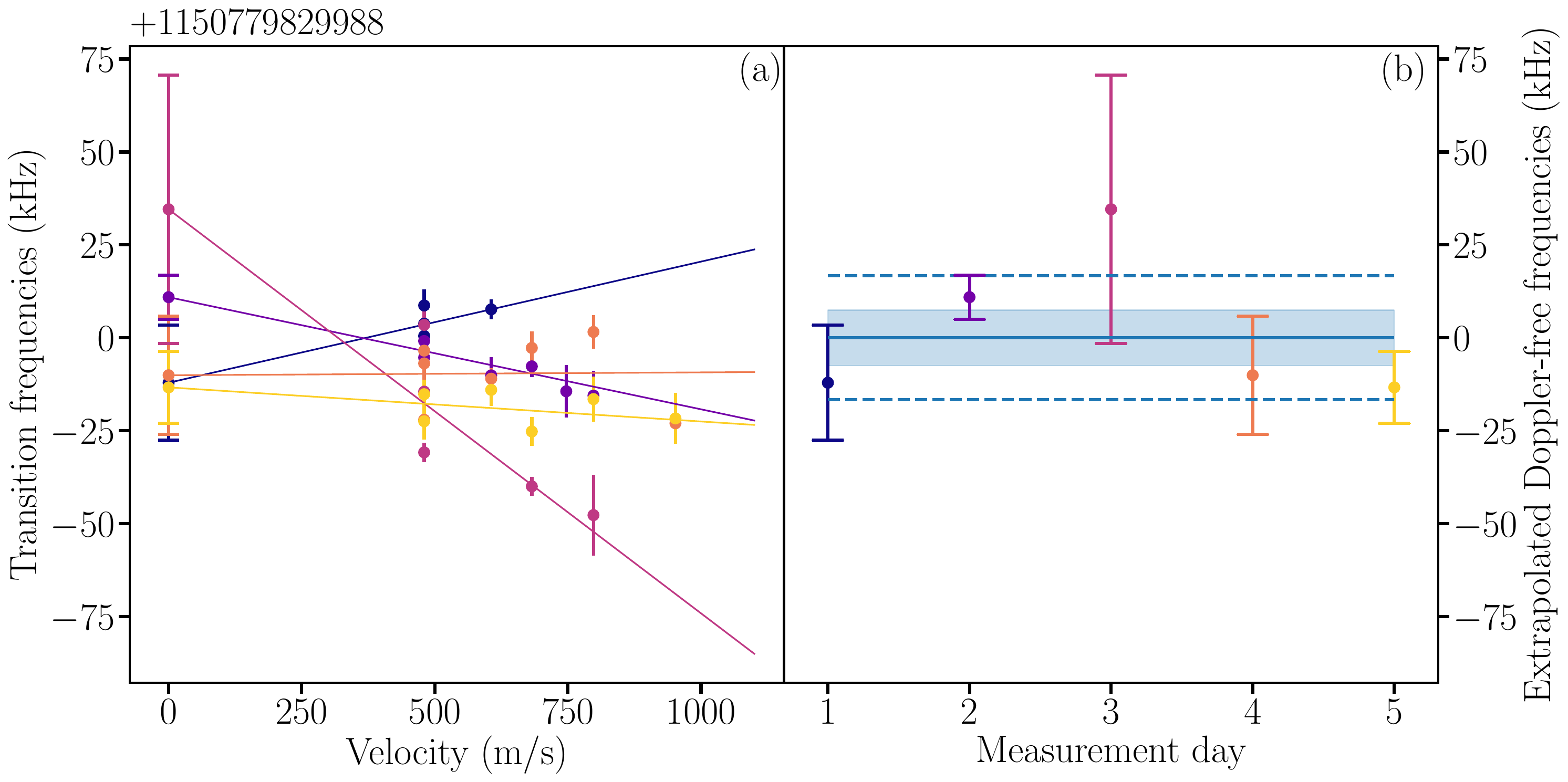} 
\caption{(a) Centroid transition frequencies of the $(1s)(40p)\,^3P_J\,\leftarrow\,(1s)(2s)\,^3S_1$ transition measured at different atomic-beam velocities. Data points of the same color correspond to measurements taken for a given alignment. Solid lines represent the least-squares extrapolation to zero beam velocity. (b) Distribution of the extrapolated Doppler-free transition frequencies with the corresponding uncertainties. The solid line shows the weighted average, dashed lines mark the standard deviation, and the blue shading represents the standard error of the mean. \label{fig:7}}
\end{figure}

As explained in Sec. \ref{sec:doppler}, residual Doppler shifts after the interferometric alignment of the incoming and reflected laser beams can be compensated by measuring spectra of each transition for different beam velocities and extrapolating the center positions to zero beam velocity. The results of a typical set of measurements of the $(1s)(40p)\,^3P_J\,\leftarrow\,(1s)(2s)\,^3S_1$ transition are presented in Fig. \ref{fig:7}(a), where the centroid frequencies are plotted as a function of the mean beam velocity for different laser-beam alignments, as indicated by the color code. Linear extrapolation to zero beam velocity yields the Doppler-free centroid positions (colored dots with vertical error bars).
Figure \ref{fig:7}(b) displays the extrapolated centroid frequencies and their weighted average (solid line). The dotted line represents the standard deviation, and the shaded area indicates the standard deviation of the mean. At each $n$ value, a total of about 30--50 spectra were recorded corresponding to at least 5 independent laser alignments. A statistical uncertainty of $7.5\,\mathrm{kHz}$ in the centroid frequency of the $(1s)(40p)\,^3P_J\,\leftarrow\,(1s)(2s)\,^3S_1$ transition was achieved in this procedure, and similar uncertainties were obtained for the other transitions investigated in this work, see Table \ref{tab:3}. 
The typical spread of the frequency residuals from the average centroid transition frequency is depicted in the inset of Fig. \ref{fig:8} (below) for all measurements of the $(1s)(50p)\,^3P_J\,\leftarrow\,(1s)(2s)\,^3S_1$ transition, with the standard deviation indicated by the dashed lines.
\subsection{Systematic uncertainties}
The transition frequencies determined as explained above need to be corrected for several systematic shifts. Table \ref{tab:1} provides a summary of the systematic shifts considered in this work with their estimated uncertainties. 
By far the dominant correction stems from the absorption-induced photon recoil, which shifts the transition frequencies to the blue by $h\nu_n/2mc^2$. This shift varies from 731.3 kHz for the transitions to the $27p$ Rydberg states to 735.7 kHz for the transitions to the $55p$ Rydberg states, with negligible uncertainties.
\\ \indent 
 \begin{figure}[t]
	\includegraphics[width=0.7\columnwidth]{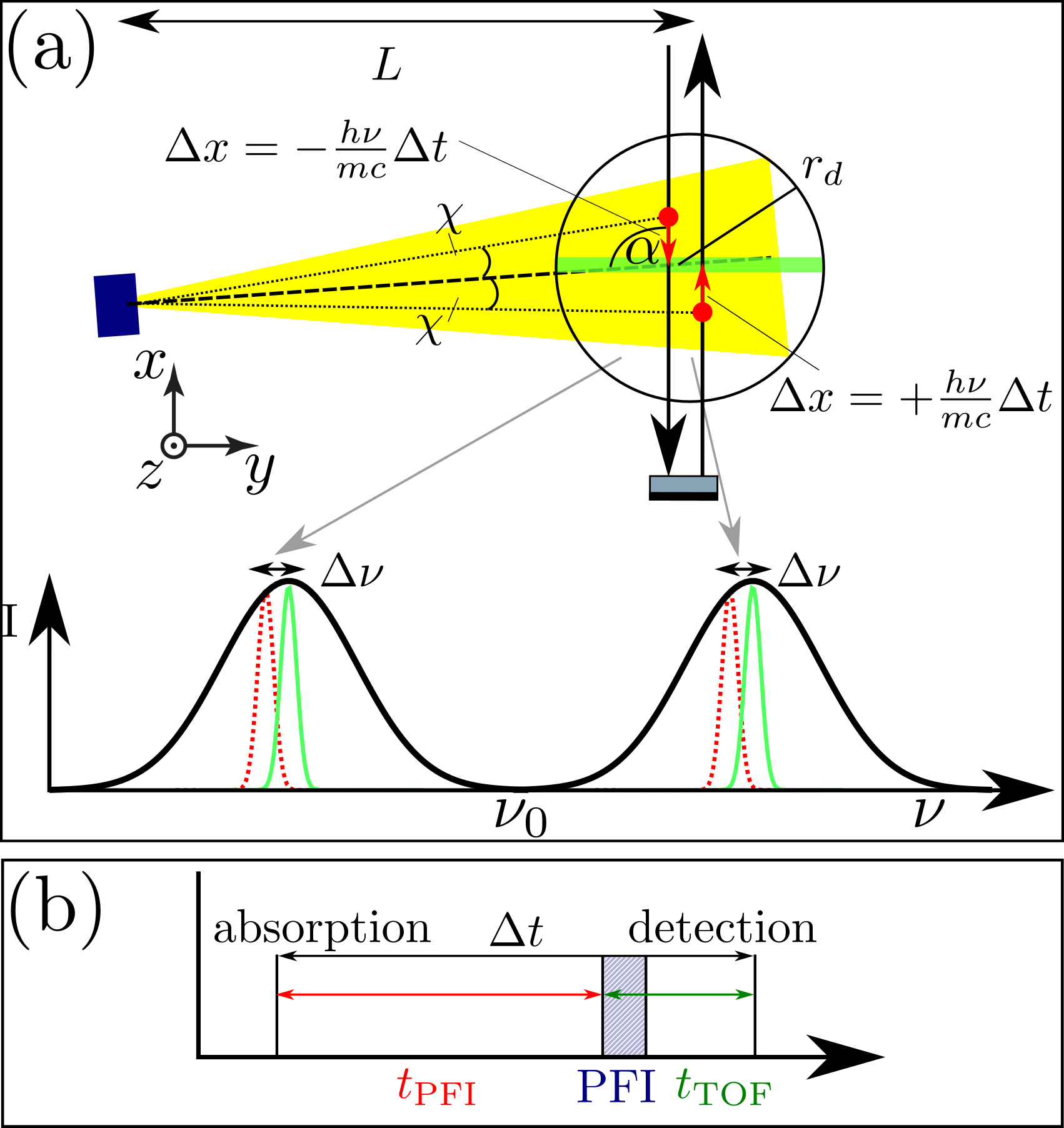}  
	\caption{Origin of the post-selection Doppler shift. (a) Top: Geometric arrangement with the relevant angles $\alpha$ and $\chi$. Red arrows designate the detection displacement caused by the photon-absorption-induced momentum transfer. The green bar indicates a selected detection stripe and the circle (radius $r_d$) marks the outer edge of the detection area. Bottom: Illustration of the post-selection red shift $\Delta \nu$. (b) Experiment time sequence. See text for details. 
		\label{fig:post_selection}}
\end{figure}
Because of the momentum transfer $\Delta p=h/\lambda_\mathrm{UV}$ upon photoabsorption, the ions detected at a given position on the detector surface that were excited from the forward- and the backward-propagating laser beam, stem from different transverse-velocity classes of the atomic beam and thus have a different velocity component along the laser-beam propagation axes [see red arrows in Fig. \ref{fig:post_selection}(a)]. Consequently, the ionized atoms have a different Doppler shift than implied by their impact position on the detector. This effect (referred to as post-selection Doppler shift hereafter), which has been observed and analyzed in a similar situation by \citet{sun24a}, is illustrated in Fig. \ref{fig:post_selection}(a) for atoms detected in the detector stripe highlighted by the green line. Absorption from the incoming and reflected laser beams (black arrows) induces a velocity change $\frac{\Delta p}{m}=\frac{h\nu_\mathrm{UV}}{mc}$ in the respective laser-beam-propagation directions and thus leads to a shift $\Delta x=\pm \frac{h\nu_\mathrm{UV}}{mc} \Delta t$ of the detected positions, where $\Delta t$ represents the time between photoexcitation and detection (the $-$ and $+$ signs correspond to the incoming and reflected beams, respectively). \\ \indent In our experiments, this time is the sum of the delay time $t_\mathrm{PFI}$ between absorption and the pulsed-field ionization (PFI) and the time of flight (TOF) $t_\mathrm{TOF}\simeq1\,\mu\mathrm{s}$ of the He$^+$ ions to the detector [see Fig. \ref{fig:post_selection} (b)]. 
Because photoexcitation is induced by a cw laser, the average value of $t_\mathrm{PFI}$, $\langle t_\mathrm{PFI} \rangle$, corresponds to half of the transit time of the atoms from the photoexcitation spot to the outer edge of the detection area [$r_d$ in Fig. \ref{fig:post_selection}(a)], which is given by the $4$-mm radius of the hole in the mumetal shield in $z$-axis TOF separating the photoexcitation and detection regions. At a beam velocity of $500\,\mathrm{m/s}$, $\langle t_\mathrm{PFI} \rangle$ is equal to $4\,\mu\mathrm{s}$, so that $\langle \Delta t \rangle=5\,\mu\mathrm{s}$. During this time, the momentum transfer leads to an average displacement of the impact position on the detector by $\langle \Delta x \rangle=\pm \frac{h\nu_\mathrm{UV}}{mc} \langle \Delta t \rangle=\pm1.9\,\mu\mathrm{m}$, which corresponds to an additional angle $\langle \chi \rangle=\arctan \left(\frac{\langle \Delta x \rangle}{L}\right )$, see Fig. \ref{fig:post_selection}(a). Consequently, the transition frequencies of atoms having absorbed a photon from both the incoming and reflected beams have a corresponding additional post-selection red shift $\langle \Delta \nu \rangle = \nu_\mathrm{UV} \frac{v}{c}\sin \langle \chi \rangle=-2\,\mathrm{kHz}$, which needs to be corrected for. In addition, we include a systematic uncertainty of $400$ Hz to account for the fact that the laser beams might not traverse the photoexcitation region centrally. This systematic uncertainty corresponds to a displacement of the laser-propagation axis by $1\,\mathrm{mm}$. 
\\ \indent 
Residual magnetic fields can cause Zeeman shifts in triplet states. The double-layer magnetic shield surrounding the photoexcitation region reduces magnetic stray fields to below $1\,\mathrm{mG}$. Because the generation of an unbalanced $m_J$ population by optical pumping can be ruled out in our experimental configuration (linear laser polarization), we do not expect any systematic error from Zeeman shifts but take a conservative systematic uncertainty of $100\,\mathrm{Hz}$ for a possible imbalance in the $m_J$ distribution of the He$^*$ atoms generated in the discharge.
\\ \indent 
The scattering of the Rydberg electron by atoms located within its orbit is the main origin of pressure shifts in Rydberg states \citep{fermi34a, amaldi34a}. Because the Rydberg electron density scales as $\frac{1}{V}$ (V is the volume) and the number of atoms responsible for the pressure shift scales as $V$, the pressure shift does not, in first approximation, depend on $n$. In our experiments, the dominant species in the photoexcitation region are ground-state He atoms, which have a pressure-shift coefficient of $5.75\,\mathrm{kHz}/(10^{12}/\mathrm{cm}^3)$ \citep{amaldi34a, koehler87a}. To estimate the pressure shift in our experiments, we use a fast ionization gauge to obtain an upper bound of the He density of $\sim 2\cdot 10^{10}/\mathrm{cm}^3$ in the photoexcitation region, which would result in a pressure shift of 100 Hz. We include a corresponding systematic uncertainty of 100 Hz. 
\\ \indent 
\begin{figure}[ht]
\includegraphics[width=\columnwidth]{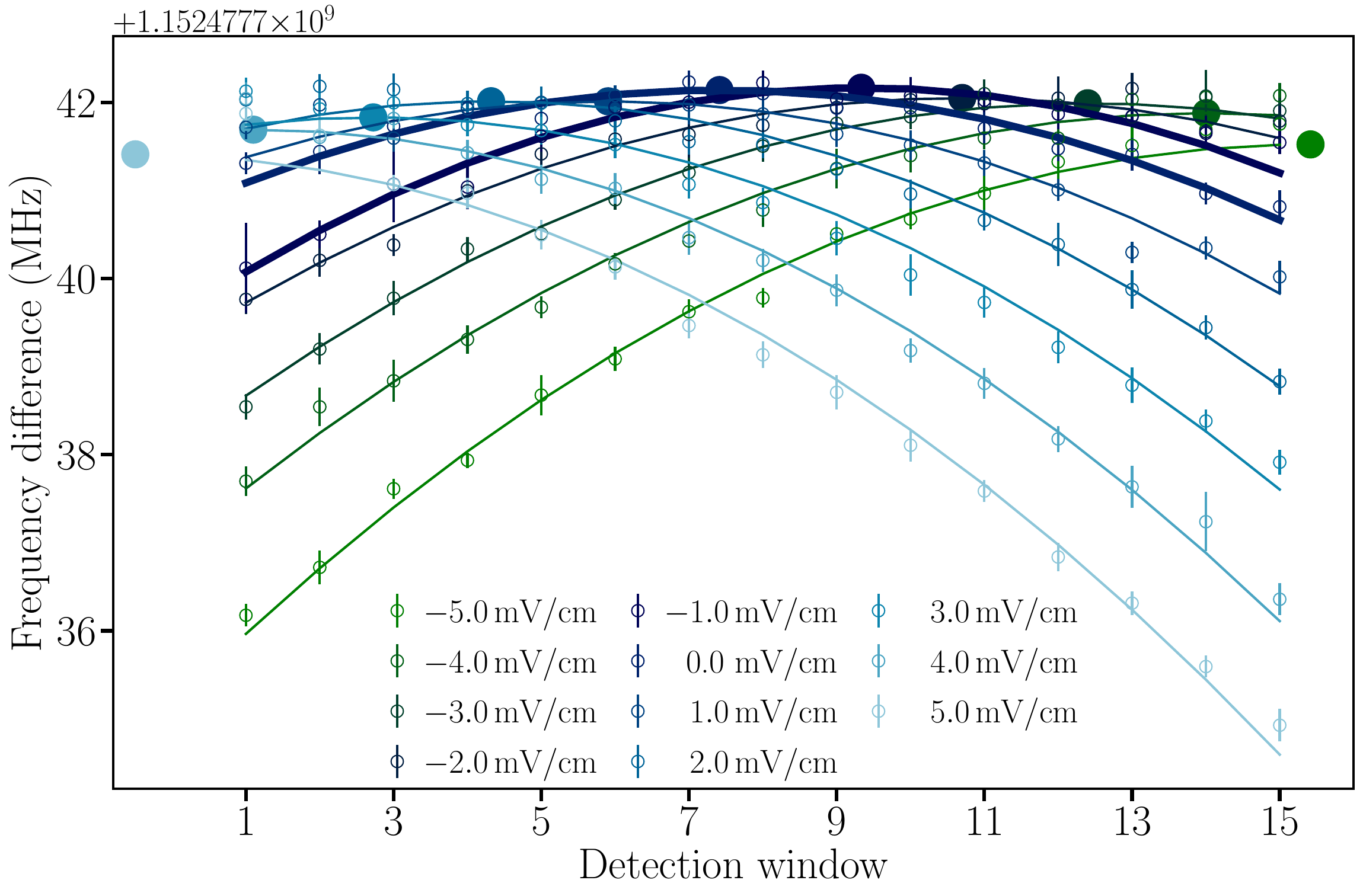}
\caption{Illustration of the electric-field compensation in $x$ direction with the example of the $(1s)(100p)\,^3P_J\,\leftarrow\,(1s)(2s)\,^3S_1$ transition of He. The open circles represent the frequencies measured in the different stripes (windows) of the imaging detector for the different compensation potentials, as indicated by the color code and legend. The full lines are the corresponding fitted parabola and their apices are marked by full circles. See text for details. \label{fig:field}}
\end{figure}
To minimize the Stark shifts from stray electric fields and the related systematic uncertainties, these fields are compensated along the three spatial dimensions by applying electric potentials on three pairs of electrodes located on opposite sides of the photoexcitation region, following the procedures established in Refs. \onlinecite{osterwalder99a, beyer18a} (see in particular Fig. 3 of Ref. \onlinecite{beyer18a}). The compensation field applied along the $z$ axis is homogeneous because the corresponding electrodes are planar. In contrast, narrow cylindrical electrodes are used to apply compensation fields in the $x$ and $y$ directions, which leads to inhomogeneous distributions along the $x$ and $y$ axes, with full field compensation being achieved only near the center of the photoexcitation volume. The photoexcitation region is a 8-mm-long, 2-mm-diameter cylinder with axis oriented along the $x$ direction. Consequently, the field inhomogeneities only lead to significant Stark shifts in this direction.   
\\ \indent We measure the field distribution $F(x)$ along the $x$ axis by recording imaging-assisted spectra of transitions to very high Rydberg states ($n\geq 100$) and determining the position-dependent transition frequencies, which are affected by a quadratic Stark shift \begin{equation}
\Delta \nu_n(x)=-\frac{1}{2} \alpha_n F^2(x),
\end{equation} where $\alpha_n=\alpha^*\cdot n^7$ is the polarizability and $\alpha^*$ is the $n$-independent polarizability coefficient of the triplet $np$ series. The measurement yields a value $\alpha^*\approx 3.15\cdot 10^{-15}\,\mathrm{MHz/(mV/cm)^2}$, see also Ref. \onlinecite{brown24a}. The results of a typical set of measurements are presented in Fig. \ref{fig:field}, which shows that the centroid frequencies of the transitions to the $100p\;^3P_J$ states recorded in the different detector windows form a distinct parabolic curve along the $x$ axis for each specific compensation potential difference (labeled by color) applied to the corresponding electrode pair. The compensated-field positions correspond to the apex of the parabolas (marked as full circles in Fig. \ref{fig:field}) and shift along the $x$ axis with increasing applied field, as expected. They also slightly shift in frequency because of the remaining field inhomogeneity resulting from our electrode geometry; it is indeed only in the center of the photoexcitation volume that the field can be exactly compensated in all three directions. 
\\ \indent The two curves displayed as bold lines in Fig. \ref{fig:field} have the apex closest to the center of the photoexcitation region (corresponding to the eighth detection window) and are used to interpolate the optimal compensation potential. At $n=100$, the field inhomogeneity induces Stark shifts of up to 1 MHz at the edges of the photoexcitation region, with an average value of $\sim 500(50)\,\mathrm{kHz}$ over the entire volume. Based on the $n^7$ and $F^2$ scaling laws of the Stark shifts, these results can be used to accurately determine the Stark shifts of transitions to any $np$ Rydberg state and their uncertainties. For instance, the systematic shifts at $n=27$ and $n=55$ are $-0.05(1)$ kHz and $-7.6(1.5)$ kHz, respectively. They are much less than the $\sim1$ MHz linewidth but need to be corrected for. A contribution of 600 Hz to the systematic uncertainty of the ionization energy results from the uncertainties of the shifts at the different $n$ values.
\\ \indent
The $2^\mathrm{nd}$-order Doppler shift, expressed as $-(\nu_n v^2/2c^2)$, results in a red shift of the transition frequencies. The forward beam velocity and its uncertainty were determined in separate time-of-flight measurements, resulting in shifts between $-7.5(5)\,\mathrm{kHz}$ and $-1.5(1)\,\mathrm{kHz}$ for beam velocities ranging from $1080(40)\,\mathrm{m/s}$ to $480(15)\,\mathrm{m/s}$, respectively. Most measurements were performed at low forward velocities and the overall uncertainty for the second-order Doppler shift is estimated to be $0.2\,\mathrm{kHz}$.
\\ \indent
Thermal blackbody radiation induces an ac-Stark shift of the transition frequencies to Rydberg states \citep{cooke80a,farley81a} which, in the $n$ range of interest, amounts to $2.2\,\mathrm{kHz}$ and must be corrected for. Because the thermal radiation in the photoexcitation volume may deviate from that of a pure blackbody, we include a systematic uncertainty of $1.1\,\mathrm{kHz}$. 
\\ \indent Quantum-interference effects often introduce systematic uncertainties in experiments with angular-dependent detection \citep{udem19a}. Our experiments are not sensitive to these effects because the pulsed-field ionization has a $100\,\%$ efficiency and acts uniformly across the entire $4\pi$ solid angle. \\ \indent
The counter-propagating geometry of the laser beams creates a standing wave, which could potentially modulate the spatial profile of the atomic beam and affect the spectral lineshape, leading to a laser-power-dependent light-force shift \citep{zheng19a}. To investigate this effect, we varied the laser power by a factor of 4 while keeping all other parameters and the laser alignment constant. No significant trend was observed within the measurement uncertainty. We attribute the absence of laser-power dependence to the weak transition moment to Rydberg states and their inherently extremely narrow natural linewidth. 
\begin{table}[h] \caption{Overview of systematic shifts and uncertainties for the centroids of the measured $(1s)(np)\,^3P_J\,\leftarrow\,(1s)(2s)\,^3S_1$ transitions of $^4$He. All values in kHz.\label{tab:1}}
\begin{tabular}{l@{\hspace{10pt}}lll}
\hline \hline Source & $\Delta \nu$& $\sigma_\mathrm{stat}$ & $\sigma_\mathrm{sys}$\\ \hline 
$1^\mathrm{st}$-order Doppler shift \footnotemark[1]&0&$7.5$ to $17$&0\\
Photon-recoil shift \footnotemark[2] &731.3 to 735.7&0 &0 \\
Post-selection shift \footnotemark[3] &$-2.5$ to $-2$ &0&0.4 \\ac-Stark shift &$2.2$&&$1.1$ \\
Zeeman shift &0& 0&0.1\\
Pressure shift &0 &0& $0.1$ \\
dc-Stark shift ($n=55$) \footnotemark[2]&0 to $-7.6$ &0 &$0.6$\\
$2^\mathrm{nd}$-order Doppler shift \footnotemark[3] &$-7.5$ to $-1.5$ &0&0.2 \\ \hline
Sum & n.a.& n.a. & 2.5 \footnotemark[4]  
\\ \hline \hline
\footnotetext[1]{depends on the number of independent measurement sets}
\footnotetext[2]{depends on the principal quantum number; extreme values for $n=27$ and 55}
\footnotetext[3]{depends on the atomic-beam velocity; extreme values for 1080 and $480\,\mathrm{m/s}$, respectively}
\footnotetext[4]{assuming that the systematic uncertainties add up linearly}
\end{tabular}
\end{table}
\begin{table}[h] \caption{$n$-dependent corrections of the photon recoil $\Delta \nu_n^\mathrm{r} $ and the dc-Stark shifts $\Delta \nu_n^\mathrm{S} $, and corrected $^4$He $(1s)(np)\,^3P_J\,\leftarrow\,(1s)(2s)\,^3S_1$ centroid transition frequencies $\nu_n$ with statistical uncertainties ($1\sigma$) given in parentheses. The last column gives the residuals of the fits of the Rydberg-Ritz formula to the observed transition frequencies. All frequencies are given in kHz.  \label{tab:3}}
\begin{tabular}{l@{\hspace{6pt}}c@{\hspace{5pt}}c@{\hspace{5pt}}l@{\hspace{0pt}}r}
\hline \hline 
$n$&$\Delta \nu_n^\mathrm{r} $& $\Delta \nu_n^\mathrm{S} $ & $\nu_n$ (corrected)&$\nu_\mathrm{exp.}-\nu_\mathrm{calc.}$ \\&(recoil)&(dc-Stark)&& \\ \hline  
27 & $-731.3$& 0.0& $1\,148\,307\,621\,273.8(10.6)$&$-0.1$ \\ 
29 & $-732.1$& 0.1&  $1\,148\,912\,959\,548.8(17.0)$&12.3 \\
33 & $-733.2$& 0.2& $1\,149\,809\,632\,366.6(12.9)$&$-4.1$\\
35 & $-733.7$& 0.3& $1\,150\,147\,008\,566.8(8.0)$&$-10.9$\\
40 & $-734.5$& 0.8& $1\,150\,779\,829\,986.7(7.5)$&7.7\\
50 & $-735.4$& 3.9& $1\,151\,523\,381\,730.5(9.8)$&$-1.8 $\\
55 & $-735.7$& 7.6& $1\,151\,752\,633\,012.0(14.1)$&$6.9$
 \\ \hline \hline
\end{tabular}
\end{table}
\subsection{Rydberg-series extrapolation}
The ionization energy $E_\mathrm{I}(2\,^3S_1)$ of the $(1s)(2s)\,^3S_1$ state is obtained from the centroid Doppler-free transition frequencies $E_n/h$ in a least-squares-fit procedure based on the Rydberg-Ritz formula \citep{ritz03a}
\begin{equation}
E_n/h=E_\mathrm{I}(2\,^3S_1)/h-\frac{R_\mathrm{He}\cdot c}{(n^*)^2}, \label{eq:rydberg}
\end{equation} where
$R_\mathrm{He}=R_\infty \frac{\mu_\mathrm{He}}{m_\mathrm{e}}=109\,722.275\,548\,36(21)\,\mathrm{cm}^{-1}$ is the mass-corrected Rydberg constant and the reduced mass $\mu_\mathrm{He}$ is given by \begin{equation}
\frac{m_{\mathrm{e}} m_{\mathrm{He}^+}}{m_{\mathrm{e}}+ m_{\mathrm{He}^+}}. \end{equation} The effective principal quantum number $n^*$ is defined as $n-\delta(n)$, where \begin{align} 
\delta(n)=&\delta_0+\frac{\delta_2}{[n-\delta(n)]^2} 
\label{eq:quantum_defect} \\&+\frac{\delta_4}{[n-\delta(n)]^4}+\frac{\delta_6}{[n-\delta(n)]^6}+\frac{\delta_8}{[n-\delta(n)]^8}+... \nonumber
\end{align} is the $n$-dependent quantum defect of the centroid of the $np\;^3\mathrm{P}$ states.
In Eq. \ref{eq:quantum_defect}, the terms proportional to $\delta_{2,...,8}$ describe the energy dependence of the quantum defect. These parameters are deduced in a Rydberg-series fit using the data presented in Table \ref{tab:3}, the $2\,^3P_J\,\leftarrow\,2\,^3S_1$ centroid frequency reported by \citet{canciopastor04a} and the calculated ionization energies of low-lying $np$ Rydberg states ($n<10$), obtained in first-principles calculations including relativistic and QED corrections, as reported in Refs. \onlinecite{drake99a, morton06a}. The procedure is described in detail in the Appendix. 
 
\begin{figure}[hb]
\includegraphics[width=\linewidth]{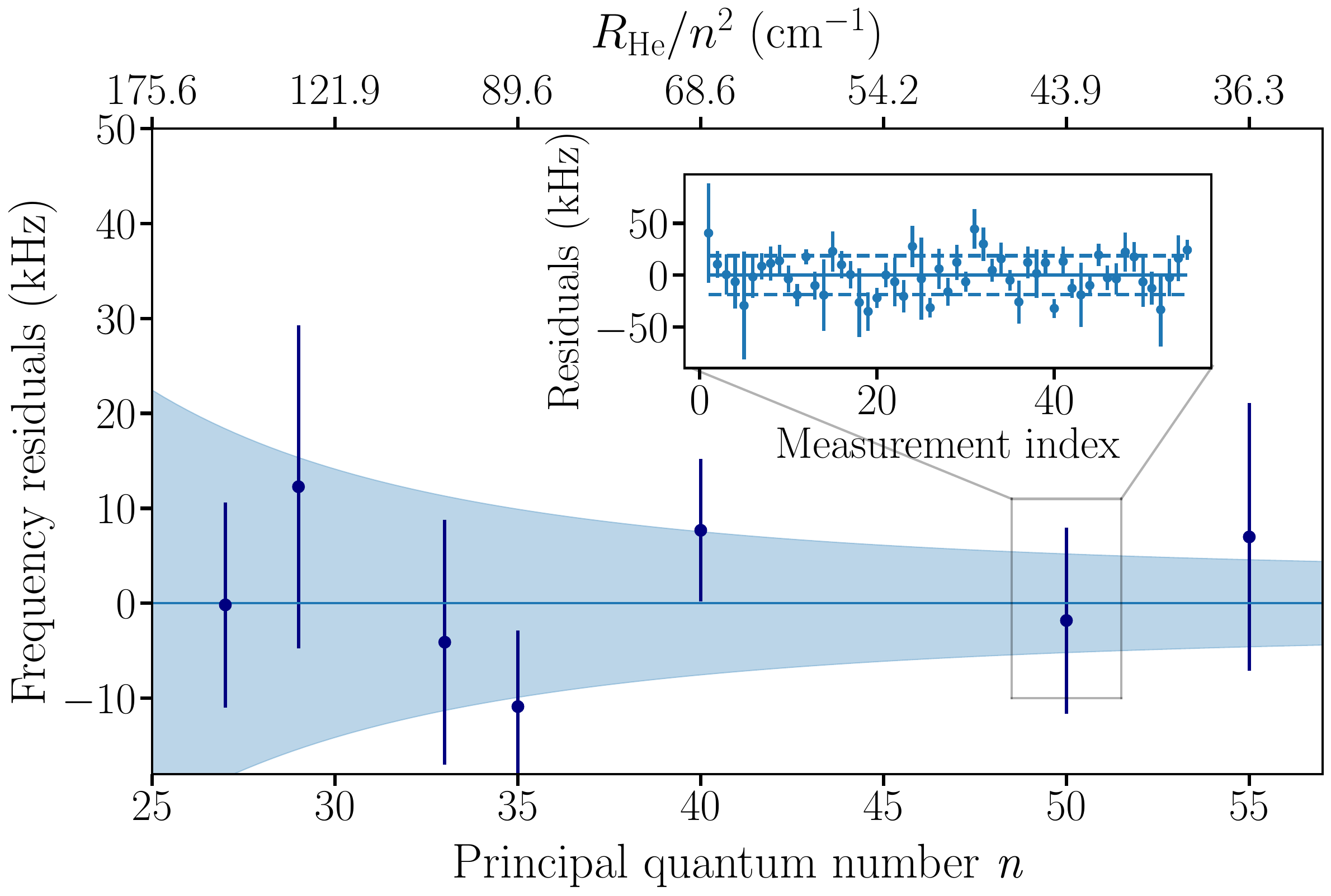} 
\caption{Frequency residuals of the $np$ series of triplet He obtained in a least-squares fit based on the Rydberg-Ritz formula (Eqs. \ref{eq:rydberg} and \ref{eq:quantum_defect}). The error bars for individual data points represent the standard deviation and the top axis indicates the binding energies ($\frac{R_\mathrm{He}}{n^2}hc$) of the corresponding Rydberg states. The blue-shaded region represents the fit uncertainty of the Rydberg-series extrapolation. The inset displays the distribution of the $(1s)(50p)\,^3P_J\,\leftarrow\,(1s)(2s)\,^3S_1$ centroid transition frequencies obtained in 55 measurements and five different laser alignments.  \label{fig:8}}
\end{figure}

The fit results are summarized in Table \ref{tab:4} and the residuals $\nu_\mathrm{exp.}-\nu_\mathrm{calc.}$ for $n=27$ to 55 are displayed in Fig. \ref{fig:8} and listed in the last column of Table \ref{tab:3}. The residuals for the low-$n$ values are provided in Table \ref{tab:app}. The new value of the ionization frequency of He$^*$ is $E_\mathrm{I}(2\,^3S_1)/h= 1\,152\,842\,742.7082(55)_\mathrm{stat}(25)_\mathrm{sys}\,\mathrm{MHz}$ and its statistical uncertainty is dominated by the contribution of the $1^\mathrm{st}$-order Doppler shift. The dominant contribution to the systematic uncertainty originates from the blackbody-radiation-induced ac-Stark shift.  \\ \indent 
\begin{table}[h] \caption{Parameters used for the Rydberg-series extrapolation  \label{tab:4}}
\begin{tabular}{l@{\hspace{10pt}}l}
\hline \hline 
Fixed parameters &Value \\ \hline 
$R_\mathrm{He}$ \footnotemark[1]& $109\,722.275\,548\,36(21)\,\mathrm{cm}^{-1}$ \\ $\delta_2(\mathrm{centroid})$ \footnotemark[2]&$-0.018\,752(4)$\\
$\delta_4(\mathrm{centroid})$ \footnotemark[2]&$ -0.011\,0(1)$\\
$\delta_6(\mathrm{centroid})$ \footnotemark[2]&$-0.015\,8(9)$\\
$\delta_8(\mathrm{centroid})$ \footnotemark[2]&$0.010(2)$\\ \hline
Fitted parameters & Value\\
 \hline 
 $E_\mathrm{I}(2\,^3S_1)/h$ \footnotemark[3]& $1\,152\,842\,742.7082(55)_\mathrm{stat}(25)_\mathrm{sys}\,\mathrm{MHz}$ \\ 
 $\delta_0(\mathrm{centroid})$ \footnotemark[3]&$0.068\, 355\, 26(4)$ \\
 \hline \hline
\end{tabular}
\footnotetext[1]{Determined using values for $R_\infty$ and electron and He$^+$ masses from \citet{ tiesinga21a}}
\footnotetext[2]{Determined using the data reported by \citet{canciopastor04a} and \citet{morton06a}, as explained in the appendix}
\footnotetext[3]{This work}
\end{table}
Our new result of the ionization energy of the $2\,^3\mathrm{S}_1$ state is compared with earlier theoretical and experimental results in Table \ref{tab:2}. It lies within $2\sigma$ ($67(33)$ kHz) of the value obtained by combining the ionization energy of the $2\,^1\mathrm{S}_0$ state \citep{clausen21a} with the $2\,^1\mathrm{S}_0 \,\leftarrow\,2\,^3\mathrm{S}_1$ transition frequency reported by \citet {rengelink18a}. The discrepancy, already pointed out in Ref. \onlinecite{clausen21a}, to the most recent theoretical value of $1\,152\,842\,742.231(52)\;\mathrm{MHz}$ \citep{patkos21a} is thus confirmed but is now at the level of $9\sigma$ $(476(54)\,\mathrm{kHz}$) because of the reduced experimental uncertainty. This discrepancy prevents the use of the calculations to determine the root-mean-square radius of the $\alpha$ particle by comparison with the experiment. Recent theoretical work has confirmed and improved important aspects of the calculations without identifying an obvious explanation for the discrepancy \citep{yerokhin22a, yerokhin23a}. \\ \indent The new values of the quantum-defect parameters of the $np$ series listed in Table \ref{tab:4} can also be used to determine the ionization energy of He$^*$ from the transition frequencies to the $33p$ and $40p$ Rydberg states reported in Ref. \onlinecite{clausen23a}. Values of $1\,152\,842\,742.688(112)\;\mathrm{MHz}$  and $1\,152\,842\,742.723(174)\;\mathrm{MHz}$ are obtained, respectively, in excellent agreement with the present results. 
 \\ \indent In future work, it will be important to extend the current investigations to $^3$He. Indeed, possible errors in the calculations are expected to cancel out in the isotopic shift, so that one could use the isotopic shift of the ionization energy of He to determine the difference of the root-mean-square charge radii of $^3$He$^{2+}$ and $^4$He$^{2+}$ \citep{pachucki15b}. The values for this difference determined from spectroscopic experiments on ``nonexotic" He atoms \citep{vanderwerf23a} and muonic He ions \citep{schuhmann23a} are in disagreement, although part of the discrepancy may be explained by an improved theoretical treatment of the singlet-triplet mixing \citep{qi24a, pachucki24a}. 
\begin{table}[h] \caption{Comparison of experimental and theoretical values of the ionization energy of the $(2\,^3S_1)$ state in He. All values are given in MHz.  \label{tab:2}}
\begin{tabular}{l@{\hspace{29pt}}l}
\hline \hline 
& $E_\mathrm{I}(2\,^3S_1)/h$ \\ \hline 
This measurement &$ 1\,152\,842\,742.7082(55)_\mathrm{stat}(25)_\mathrm{sys}$\\
\citet{patkos21a} &$1\,152\,842\,742.231(52)$ \\ \citet{clausen21a} \footnotemark[1]& $1\,152\,842\,742.640(32)$ \\ \hline \hline
\end{tabular}
\footnotetext{Obtained by adding the frequency of the $(1s)(2s)\,^1S_0\,\leftarrow\,(1s)(2s)\,^3S_1$ transition reported by \citet{rengelink18a} to the value of the ionization frequency $E_\mathrm{I}\;(2\;^1S_0)/h$ of the $(1s)(2s)\;^1S_0$ metastable state }
\end{table}
\section{Conclusions}
In this article, we have presented in detail a new approach to determine the ionization frequency $E_\mathrm{I}(2\;^3S_1)/h$ of metastable helium by Rydberg-series extrapolation. This new approach combines (i) the use of supersonic beams with adjustable forward velocities generated by a cryogenic pulsed valve, (ii) the imaging-assisted recording of Doppler-free spectra based on an interferometric retroreflection laser-alignment procedure, (iii) the spatial mapping and compensation of stray-electric fields, and (iv) the calibration of the laser frequency using an SI-traceable frequency-distribution network. We have also detailed the procedure we developed to characterize systematic errors in the determination of Stark shifts by residual electric fields and in the cancellation of Doppler shifts by retroreflection.
 \\ \indent With this approach, the ionization frequency $E_\mathrm{I}(2\,^3S_1$) of metastable He was determined to be $1\,152\,842\,742.7082(55)_\mathrm{stat}(25)_\mathrm{sys}\,\mathrm{MHz}$, the uncertainties being dominated by the statistical compensation of the residual 1$^\mathrm{st}$-order Doppler shifts and the Rydberg-series extrapolation (5.5 kHz) and the systematic contribution from the ac-Stark shift induced by the thermal radiation (1.1 kHz). This new value represents an improvement by almost one order of magnitude over the previous most precise experimental value of the ionization energy of metastable He. It also confirms a discrepancy at the $9\,\sigma$ level with the most precise theoretical result reported so far \citep{patkos21a}. 
 \\ \indent In order to obtain additional data that may help clarifying the origin of this discrepancy we intend, as a next step, to determine the ionization energy of $^3$He using the same approach. A precise determination of the isotopic shift of the ionization energy of He would also provide an independent value of the squared-charge-radii difference $\Delta r^2$ between the $^3$He$^{2+}$ and $^4$He$^{2+}$ nuclei that may help resolving the current discrepancy between the $\Delta r^2$ values derived from the Lamb-shift measurements in muonic He$^+$ \citep{krauth21a, schuhmann23a} and from precision spectroscopy of the $(1s)(2s)\,^1S_0\,\leftarrow\,(1s)(2s)\,^3S_1$ transition \citep{rengelink18a, vanderwerf23a}. 
 \\ \indent Helion ($^3$He$^{2+}$) has an $I=1/2$ nuclear spin and the $1s \, ^2S_{1/2}$ ground state of $^3$He$^+$ is split by the hyperfine interaction into two levels with total angular momentum quantum numbers $F^+=1$ and $F^+=0$ separated by $8\,665\,649\,865.77(26)_\mathrm{stat}(1)_\mathrm{sys}\;\mathrm{Hz}$ \citep{schneider22a}. The hyperfine interaction profoundly affects the Rydberg-level structure, in particular by inducing mixing between the Rydberg series of singlet ($S=0$) and triplet ($S=1$) character. The framework needed to reliably extrapolate the $np$ Rydberg series of $^3$He, multichannel quantum defect theory, was established by \citet{vassen89a} and similar approaches were already applied to precision measurements of the hyperfine structure of other atomic \citep{schaefer06b, schaefer10a, herburger24a, peper24a} and molecular ions \citep{osterwalder04a}. We expect to be able to determine the ionization energy of $^3$He$^*$ with a similar accuracy as reported here for $^4$He$^*$, which would result in an uncertainty of less than $0.01\;\mathrm{fm}^2$ in $\Delta r^2$, i.e. less than the difference between the $\Delta r^2$ values reported in references \citep{krauth21a, schuhmann23a} and \citep{rengelink18a, vanderwerf23a}.  
\begin{acknowledgments}
We thank Krzysztof Pachucki for discussions and his encouragements and for making an early version of Ref. \onlinecite{pachucki24a} available prior to submission.
This work is supported financially by the Swiss National Science Foundation under grants No. CRSII5-183579 and No. 200020B-200478.
\end{acknowledgments}
\appendix*
\section{Quantum-defect parameters}
\label{appendix}
The measurements presented in the article concern transitions to ($1s)(np)\,^3P_J$ Rydberg states of $^4  $He with $n$ values in the range between $n=27$ and 55, and cover a range of binding energies corresponding to $3.4\times10^6$ MHz. Over this restricted range, the transition frequencies only very weakly depend on the energy-dependent terms of the quantum defect ($\delta_2$, $\delta_4$, $\delta_6$ and $\delta_8$ in Eq. \ref{eq:quantum_defect}) and therefore cannot be used to derive the values of the parameters $\delta_{2,...,8}$, which are primarily determined from the positions of the lowest members of the ($1s)(np)\,^3P_J$ Rydberg series. 

\begin{table}[h] \caption{$n$-dependent $^4$He $(1s)(np)\,^3P_J\,\leftarrow\,(1s)(2s)\,^3S_1$ centroid transition frequencies $\nu_n$ reported by \citet{canciopastor04a} and \citet{morton06a} and residuals of the Rydberg-series adjustment $\nu_n-\nu_\mathrm{fit}$. All frequencies are given in MHz.  \label{tab:app}} 
	\begin{tabular}{r@{\hspace{1.5cm}} S@{\hspace{1.5cm}} r}
		\hline \hline
		$n$ & $\nu_n$&  $\nu_n-\nu_\mathrm{fit}$ \\
		\hline
		2  & 276736495.6531(24)&0.00 \\
		3  & 770725260.0(2.6) &$-0.02$ \\
		4  & 940181611.0(2.5)  &$0.41$\\
		5  & 1017637745.5(2.5)  &$-1.76$\\
		6  & 1059369811.4(2.5)    &$0.76$\\
		7  & 1084389599.5(2.5)    &$1.49$\\
		8  & 1100560279.4(2.5)   &$1.24$\\
		9  & 1111611199.4(2.5)    &$0.72$\\
		10 & 1119495769.0(2.5)    &$0.22$\\
		\hline \hline
	\end{tabular}
\end{table}
To obtain a global description of the $np$ series, we therefore combine the results of our new measuremenets with the value of the $(1s)(2p)\,^3P_J\,\leftarrow\,(1s)(2s)\,^3S_1$ transition frequency reported by \citet{canciopastor04a} and with values of the $(1s)(3-10p)\,^3P_J\,\leftarrow\,(1s)(2s)\,^3S_1$ transition frequencies calculated by \citet{morton06a}, as listed in Table \ref{tab:app}. The accuracy of these low-$n$ data is better than 2 MHz which enables the derivation of accurate values of $\delta_{2,...,8}$. The ionization energy $E_\mathrm{I}$ of the $(1s)(2s)\,^3S_1$ state of $^4$He reported in the article is obtained in an iterative two-step fit procedure, based on Eqs. \ref{eq:rydberg} and \ref{eq:quantum_defect}. In the first step, $E_\mathrm{I}$ and $\delta_0$ are determined from the high-$n$ data (Table \ref{tab:3}). They are then kept fixed in the second step in which the values of $\delta_{2,...,8}$ are fitted using the $n=2-10$ data. The procedure is repeated until convergence is reached. The final results of the fit are provided in Table \ref{tab:4}. The residuals of all transition frequencies are presented in Table \ref{tab:3} and \ref{tab:app} and indicate that Eqs.~\ref{eq:rydberg} and~\ref{eq:quantum_defect} with the fit parameters in Table~\ref{tab:4} adequately describe all precisely known frequencies of the $(1s)(np)\,^3P_J\,\leftarrow\,(1s)(2s)\,^3S_1$ Rydberg series.

%

\end{document}